\documentclass[journal=jctcce,manuscript=article,layout=traditional]{achemso}
\usepackage{fullpage}
\usepackage{mathtools}
\usepackage[usenames,dvipsnames]{xcolor}
\usepackage{amsthm}
\usepackage{amssymb}
\usepackage{bm}
\usepackage[version=3]{mhchem}
\usepackage{braket}
\usepackage{relsize}
\usepackage{enumitem}
\usepackage[font=small,labelfont=bf]{caption}
\usepackage{graphicx}
\usepackage{footnote}
\usepackage{multirow}
\usepackage{tabularx}
\usepackage{booktabs}
\usepackage{makecell}
\usepackage{threeparttable}
\usepackage{chemfig}

\usepackage{caption}
\usepackage{subcaption}

\newcommand{\RNum}[1]{\expandafter{\romannumeral #1\relax}}

\newcolumntype{C}{>{\centering\arraybackslash}X}

\newpage
\title{Earth Mover's Distance as a metric to evaluate the extent of charge transfer in excitations using discretized real-space densities}

\author{Zhe Wang}
\altaffiliation{These authors contributed equally to this work.}
\affiliation{
	Kenneth S. Pitzer Center for Theoretical Chemistry,
	Department of Chemistry,
	University of California at Berkeley,
	Berkeley, CA 94720, USA
}
\author{Jiashu Liang}
\altaffiliation{These authors contributed equally to this work.}
\affiliation{
	Kenneth S. Pitzer Center for Theoretical Chemistry,
	Department of Chemistry,
	University of California at Berkeley,
	Berkeley, CA 94720, USA
}
\author{Martin Head-Gordon}
\affiliation{
	Kenneth S. Pitzer Center for Theoretical Chemistry,
	Department of Chemistry,
	University of California at Berkeley,
	Berkeley, CA 94720, USA
}
\alsoaffiliation{
	Chemical Sciences Division,
	Lawrence Berkeley National Laboratory,
	Berkeley, CA 94720, USA
}
\email{mhg@cchem.berkeley.edu}

\date{\today}

\begin{document}
\newpage

\begin{tocentry}
\includegraphics[width=\textwidth]{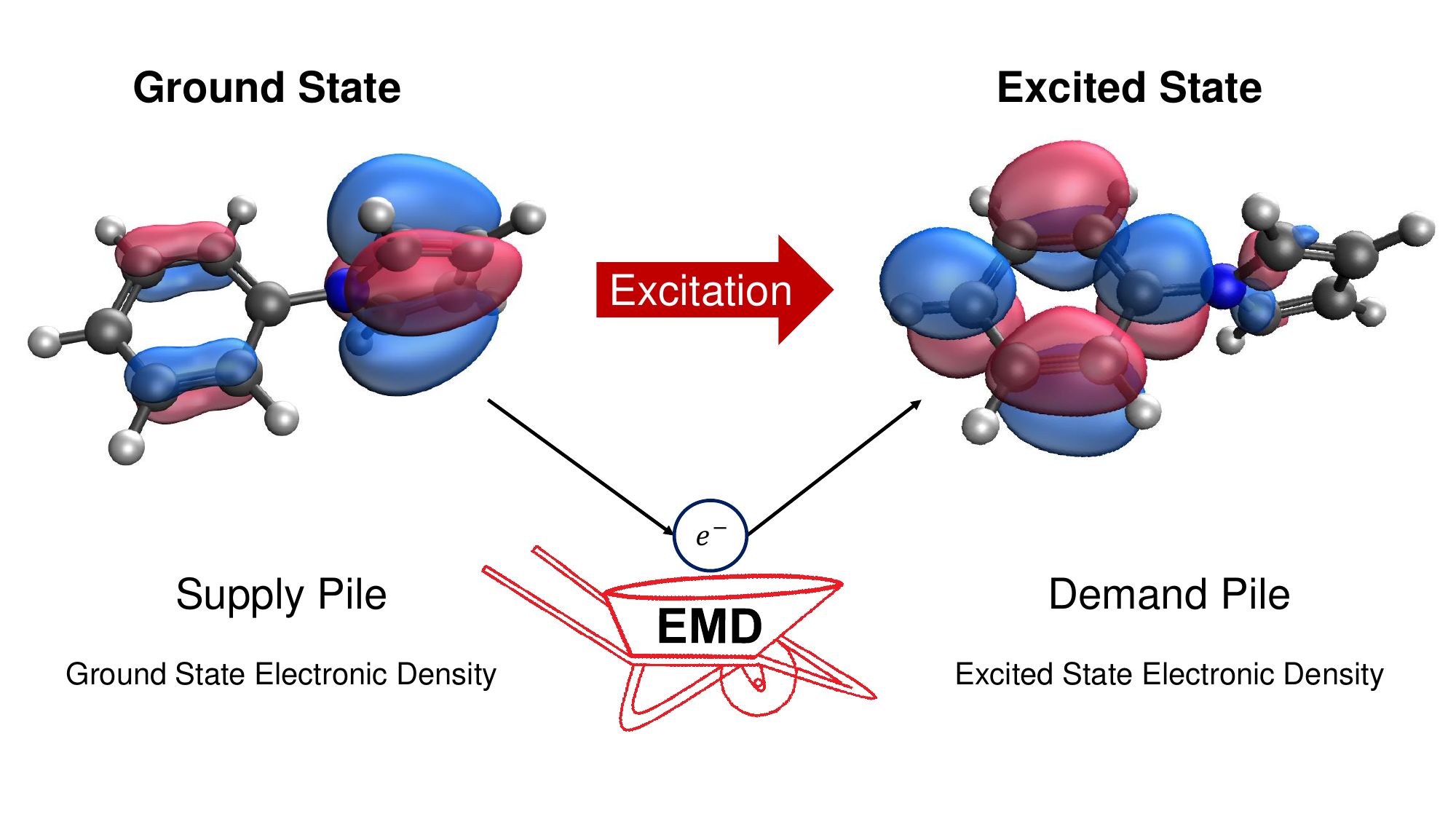}
\end{tocentry}

\begin{abstract}

This paper presents a novel theoretical measure, $\mu^{\text{EMD}}$, based on the Earth Mover's Distance, for quantifying the density shift caused by electronic excitations in molecules. As input, the EMD metric uses only the discretized ground and excited state electron densities in real space, rendering it compatible with almost all electronic structure methods used to calculate excited states. The EMD metric is compared against other popular theoretical metrics for describing the extent of electron-hole separation in a wide range of excited states (valence, Rydberg, charge-transfer, etc). The results showcase the EMD metric's effectiveness across all excitation types and suggest that it is useful as an additional tool to characterize electronic excitations. The study also reveals that $\mu^{\text{EMD}}$ can function as a promising diagnostic tool for predicting the failure of pure exchange-correlation functionals. Specifically, we show statistical relationships between the functional-driven errors, the exact exchange content within the functional, and the magnitude of $\mu^{\text{EMD}}$ values.
\end{abstract}
\maketitle
\clearpage

\section{Introduction} \label{sec:intro}
Charge transfer excitations are important in a range of photochemical applications.\cite{zhang2020delayed,heinemann2017critical,vogler2000photochemistry} It would be advantageous to assess the extent of charge transfer (CT) in a given excitation to ascertain its excitation type, thereby facilitating the discovery of novel materials such as photosensitizers that require charge transfer properties.\cite{jaeger2015using} Theoretically, the development of a metric for the extent of charge transfer may also assist in evaluating the performance of functionals in time-dependent density functional theory (TDDFT) for challenging excitations with large electron-hole separations.\cite{mester2022charge} While it may be feasible to discern the excitation types by visualizing the total density difference or the relevant molecular orbitals, this process is typically complex and becomes burdensome when investigating many excitations. Therefore, a quantitative comprehension of the charge transfer extent is essential.  

Over the past fifteen years, numerous theoretical metrics have been proposed to assess the extent of charge transfer.\cite{le2011qualitative,jacquemin2012best,campetella2019quantifying,peach2008excitation,guido2013metric,guido2014effective,head1995analysis,plasser2014new,etienne2014toward,plasser2012analysis,plasser2015statistical} These metrics typically rely on electronic density,\cite{le2011qualitative,jacquemin2012best,campetella2019quantifying} molecular orbitals,\cite{peach2008excitation,guido2013metric,guido2014effective} attachment and detachment densities,\cite{head1995analysis,plasser2014new,etienne2014toward} or the one-electron reduced transition density matrix (1-TDM)\cite{plasser2012analysis,plasser2015statistical} to predict the electron-hole distance in excitations and thereby characterize the electronic density change upon excitation. Each approach comes with its own strengths and weaknesses. For example, the $\mu^{\text{CT}}$ metric (denoted by $\mu^{\text{LBAC}}$ in this paper since it's proposed by Le Bahers, Adamo, and Ciofini) \cite{le2011qualitative} which is solely based on the electronic density in real space, is straightforward and readily adaptable to various electronic excitation calculation methods. However, it struggles to accurately characterize  the charge transfer associated with centrosymmetric excitations. Although the TDM-based methods can describe centrosymmetric excitations and have seen extensive application in TDDFT calculations, they are expected to be challenging to apply to double excitations since doubly excited configurations do not directly appear in the 1-TDM.\cite{plasser2014new} Consequently, it remains desirable to develop a new density-based metric that can characterize all types of excitations.

The Earth mover's distance (EMD), a prevalent metric in computer vision (CV) used to denote differences between distributions,\cite{rubner1998metric, levina2001earth, rubner2000earth,zhao2008differential,andoni2009efficient} could offer a solution to this challenge. In this paper, we propose a new metric called $\mu^{\text{EMD}}$ for describing the extent of charge transfer in electronic excitations based on the real-space electronic density. In subsequent sections, we will illustrate that this metric can be employed to characterize centrosymmetric excitations and is a valuable supplement to the existing metrics for electronic excitation analysis. As it relies only on the electronic density in real space (and not on any specific form of the excited state wavefunction), we will also demonstrate its ready applicability beyond the widely utilized TDDFT approaches to include a series of orbital-optimized DFT (OO-DFT) methods,\cite{hait2021orbital} which can be applied in calculating double and core excitations with substantial orbital relaxations.

We will first introduce the theory of EMD and the notations employed in this paper (Section~\ref{subsec:the model}), with a designed grid pruning strategy to reduce the computational cost of $\mu^{\text{EMD}}$ (Section~\ref{subsec:grid_pruning}). Then, we will briefly introduce some other theoretical metrics for comparison (Section~\ref{subsec:other models}) and summarize the computational details (Section~\ref{subsec:computational details}). Afterward, we will compare the performance of these theoretical metrics on excitations of different types (Section~\ref{subsec:EMD_camb3lyp}), discuss the use of $\mu^{\text{EMD}}$ in studying double and core excitations with OO-DFT methods (Section~\ref{subsec:double_ex}), demonstrate the influence of the functional choice on the metric (Section~\ref{subsec:EMDs}), and show that $\mu^{\text{EMD}}$ can be used as a diagnostic tool in identifying when the common semi-local exchange-correlation (XC) functionals will fail due to the large density changes in excitations (Section~\ref{subsec:metric_for_TDDFT}). Our conclusions are summarized in Section~\ref{sec:conclusion}.

\section{Methodology} \label{sec:method}
\subsection{EMD Model} \label{subsec:the model}

In the field of statistics, EMD serves as a measure of the difference between two probability distributions over a specified region. It is also known as the Wasserstein distance in mathematics.\cite{levina2001earth} Evaluating the EMD is an optimal transportation problem, which can be concisely described as the minimum amount of work required to move a pile of soil to a hole of identical volume. EMD is extensively employed in the field of image recognition and retrieval.\cite{rubner2000earth,zhao2008differential,andoni2009efficient} Inspired by this, we can use it to measure the difference of the charge density distributions before and after electronic excitation.

EMD can be subdivided into EMD for continuous distributions and EMD for discrete distributions depending on the continuity of the probability distribution. Monge described such problems for continuous probability distributions over two hundred years ago.\cite{monge1781memoire} Since then, Kantorovich has loosened the conditions of Monge's problem in order to solve the optimal transportation problem by finding an optimal joint distribution to minimize transportation costs.\cite{kantorovich1942translocation} However, few techniques exist for solving difficult continuous cases.\cite{chi2022approximate} Consequently, we will employ a discrete form of the Kantorovich problem.

The discrete EMD problem considers two charge distributions discretized on a grid, the ``supply pile'', $S = \{(\mathbf{r}_i, q^S_i)\}$ and the ``demand pile'', $D = \{(\mathbf{r}_j, q^D_j)\}$. 
Here $\mathbf{r}_i$ and $\mathbf{r}_j$ are the Cartesian coordinates of grid points used to discretize the two distributions respectively, and $q^S_i$ and $q^D_j$ are the weights (i.e. effective charges) associated with each grid point. To solve the EMD problem, we want to find the optimal transmission matrix $\mathbf{F} = \{f_{ij}\}$ to minimize the cost. For $\sum_{i} q_i^S = \sum_{j} q_j^D$, we have
\begin{equation}
\begin{aligned}
    & \mathbf{F} = \mathop{\arg\min}_\mathbf{F} \sum_{ij}f_{i,j}d_{i,j} \\
    & \text{s.t.} \ f_{i,j} \geq 0, \\
    & \sum_{j} f_{i,j} = q_{i}^S, \\
    & \sum_{i} f_{i,j} = q_{j}^D. 
\end{aligned}
\label{eq:lp}
\end{equation}
Here $d_{i,j}=\sqrt{(\mathbf{r}_i - \mathbf{r}_j)^2}$ is the distance between grid points $i$ and $j$. By examining the constraints, we see that the discrete optimal transportation problem is actually a linear programming problem with linear constraints.

To calculate the EMD, we select a set of quadrature grid points in three-dimensional real space, $\{\mathbf{r}_i\}$ to describe the discrete charge distributions of the ground state (GS) and excited states (ES). Using the charge density $\rho_i=\rho(\mathbf{r}_i)$ and the quadrature weight $w_i$ at the $i$-th grid point, we can obtain the total amount of charge associated with this grid point:
\begin{equation}
    q_i = w_i\rho_i.
\label{eq:qi}
\end{equation}
Applying this to the GS and ES respectively, we can get $q^\text{GS}_i$ and $q^\text{ES}_i$. We then define the following expressions for the discretized supply pile ($q^S_i$) and demand pile ($q^D_j$) associated with the electronic excitation, based on ensuring that the supply pile represents a source of electrons while the demand pile represents a sink of electrons. Note that while both grid points $i$ and $j$ originate from the identical set of quadrature grid points, distinct indices are utilized to differentiate the supply pile from the demand piles.
\begin{equation}
    \begin{aligned}
        & \Delta q_i = q_i^{\text{ES}} - q_i^{\text{GS}},\\
        & q_i^S = \left\{ \begin{array}{lr}
            0,  &  \Delta q_i > 0\\
            -\Delta q_i, & \Delta q_i \leq 0
        \end{array} \right. \\ 
        & q_j^D = \left\{ \begin{array}{lr}
           \Delta q_j,  &  \Delta q_j > 0\\
            0, & \Delta q_j \leq 0
        \end{array} \right.
    \end{aligned}
\label{eq:piles}
\end{equation}

The total transferred charge during this process can be defined as\cite{jacquemin2012best}
\begin{equation}
    q^{\text{CT}}  = \sum_i q_i^S = \sum_j q_j^D.
\label{eq:qCT}
\end{equation}
$q^{\text{CT}}$ also partly characterizes the degree of GS and ES density overlap during the excitation process. Then we optimize the transmission matrix $\mathbf{F} = \{f_{ij}\}$ to evaluate the EMD ($\mu^{\text{EMD}}$) as:
\begin{equation}
        \mu^{\text{EMD}} = \min_\mathbf{F} \sum_{ij} f_{i,j}d_{i,j}. \\
\end{equation}
Note that 
$\mu^{\text{EMD}}$ possesses units of charge $\times$ length, consistent with dipole moments. With the range of CT in mind, it is also useful to renormalize $\mu^{\text{EMD}}$ by $q^{\text{CT}}$ to define an EMD-derived distance, $d^{\text{EMD}}$, with units of length:
\begin{equation}
        d^{\text{EMD}} = \frac{\mu^{\text{EMD}}}{q^{\text{CT}}}
\end{equation}
$d^{\text{EMD}}$ represents the shortest possible distance from the ground state charge distribution to the excited state charge distribution, which is inherently non-negative. The $\mu^{\mathrm{EMD}}$ metric integrates the amount of transferred charge ($q^{\text{CT}}$) and this shortest distance, providing insight into the overall difference between the electronic densities of the excited state and the ground state. It is interesting to note that for unidirectional charge transfers in one dimension, $\mu^{\text{EMD}}$ can be reduced to the existing $\mu^{\text{LBAC}}$ metric.

\subsection{Grid Selection for Efficient EMD Calculations} \label{subsec:grid_pruning}

The simplex method, a standard algorithm for tackling linear programming problems,\cite{dantzig1990origins} is often employed to solve EMD problems. In this context, we utilize the transportation simplex algorithm, which exhibits an average polynomial time complexity indexed between 2 and 3 with respect to the number of grids. However, achieving reasonably accurate excitation energies in TDDFT often mandates an extensive number of grid points, sometimes amounting to tens of thousands for one non-hydrogen atom.\cite{gill1993standard,dasgupta2017standard} Given the intricacy of the transportation simplex algorithm, it is unrealistic to accept such a large number of grid points as input. To reduce the volume of input, we therefore introduce a smaller grid (still of the standard atom-centered type used in molecular DFT calculations) and evaluate the real-space charges on this small grid as:
\begin{equation}
    q_i = \sum_k w_k\rho_k.
\end{equation}
Here, $w_k$ and $\rho_k$ are the quadrature weight and charge density on the grid points (in the grid used for TDDFT) that will be associated with each chosen grid point $i$ (in the smaller set). All the grid points employed in the TDDFT calculation are assigned to grid points in the smaller set used for EMD according to their spatial distances. Afterward, this smaller selected set of grid points is used for the EMD calculation. By selecting the grids properly, we demonstrate that much fewer grid points are required to obtain an accurate EMD result. In this paper, the grids are chosen with the radial part treated using the Euler-Maclaurin scheme \cite{murray1993quadrature} and the angular part using the Lebedev scheme.\cite{lebedev1975values}.

\begin{figure}[ht!]
    \centering
    \includegraphics[width=0.85\textwidth]{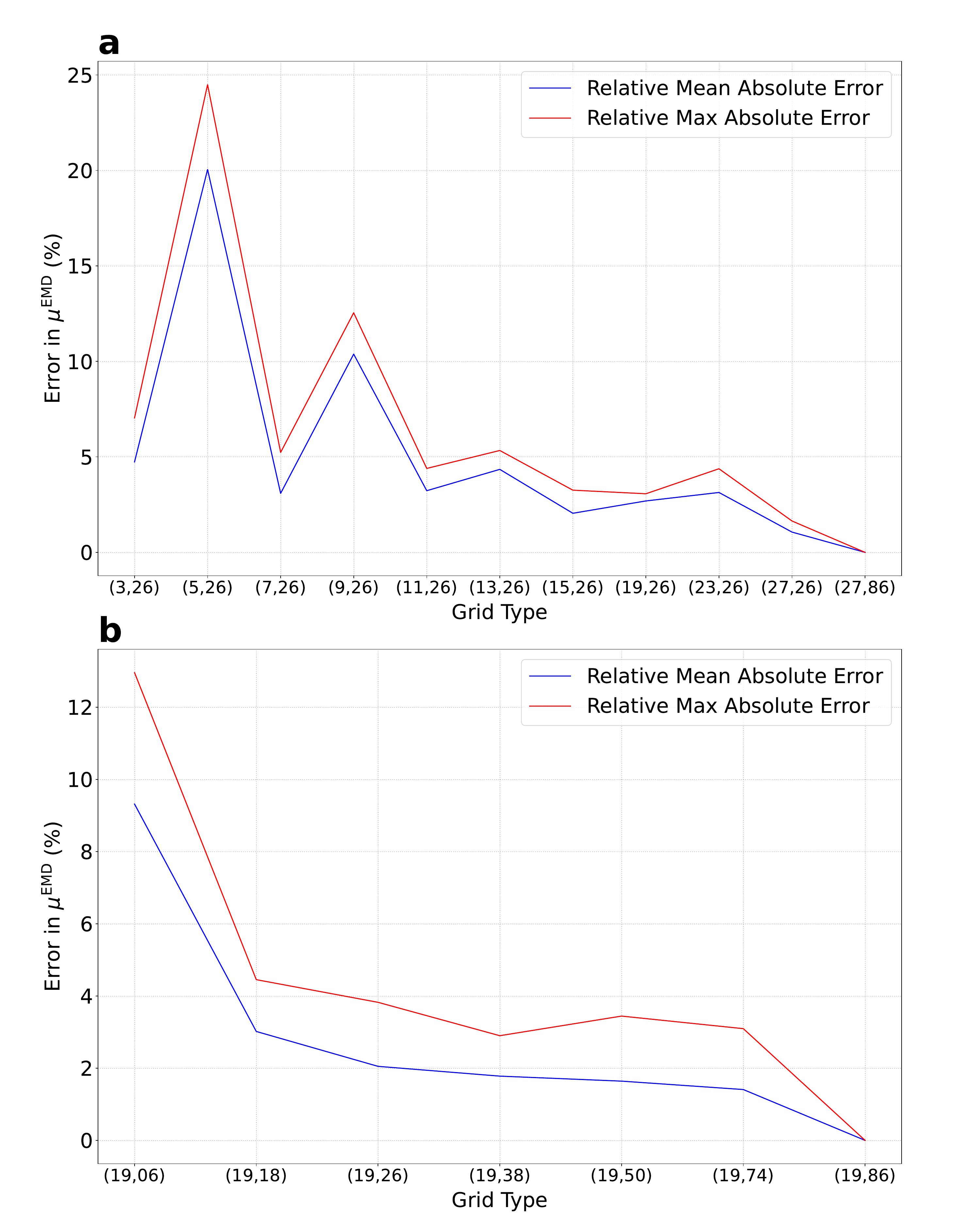}
    \caption{The convergence trend of $\mu^{\text{EMD}}$ in relation to the grid selection is displayed for challenging excited states. The x and y in "(x,y)" represent the numbers of grid points in the radial part and the angular part respectively for each atom. (a) Convergence with an increasing radial grid for three Rydberg excitations ($n \to 3s$ in water, $\pi \to 3s$ in ethylene, and $\pi \to 3s$ in furan). (b) Convergence with an increasing angular grid for five non-centrosymmetric excitations (CT excitations in HCl and Benzonitrile and Rydberg excitations in CO, water, and furan, respectively)}
    \label{fig:conv}
\end{figure}

It is expected that the convergence of the radial grid is more difficult for characterizing Rydberg states because they have larger charge variations in more diffuse regions. Therefore, three challenging cases are examined ($n \to 3s$ in water, $\pi \to 3s$ in ethylene, and $\pi \to 3s$ in furan), for which relative max absolute error (rMAX) and mean absolute error (rMAE) relative to (27,86) are displayed in Figure~\ref{fig:conv} (a). For these excited states, a minimum of 11 radial grid points per atom suffices to keep the rMAE below 5$\%$. For the angular part, more grid choice is examined for five non-centrosymmetric excitations in our data set, as they require a higher number of angular grids to depict the excitation properly. As shown in Figure~\ref{fig:conv} (b), 18 grid points are sufficient to bring the rMAE below 3$\%$. We choose (19,26) as the key grid points to generate the EMD metrics in this paper, under which circumstances the error of the most difficult case should be below 10$\%$.

After defining a suitable small grid, the computation time of $\mu^{\text{EMD}}$ is acceptable compared to TDDFT calculations. As a benchmark, TDDFT calculation takes 5.4 hours for 10 excited states of aminobenzonitrile using CAM-B3LYP/aug-cc-pVTZ/(50, 194). In contrast, EMD calculation takes 0.22 hours for a single excited state of the same molecule using the same computer and the (19, 26) grid.

\subsection{Other CT Metrics for Comparison} \label{subsec:other models}

As briefly reviewed in the Introduction, there are several widely used theoretical metrics that aim to characterize the extent of charge transfer in an electronic excitation. For reasons discussed below, we select 2 existing metrics to compare against our new EMD metric.

The first metric we will compare against is $\mu^{\text{LBAC}}$, which is based upon the real space GS and ES electron densities.\cite{le2011qualitative} $\mu^{\text{LBAC}}$ measures the change in the dipole moment between the ground and excited states,
\begin{equation}
    \mu^{\text{LBAC}} = \Big\lVert \int d^3\mathbf{r}\,  [\rho^{\text{ES}}(\mathbf{r}) - \rho^{\text{GS}}(\mathbf{r})] \,\mathbf{r} \Big \rVert,
\end{equation}
where $\rho^{\text{GS}}(\mathbf{r})$ and $\rho^{\text{ES}}(\mathbf{r})$ represent the electronic density of the ground state and excited state respectively. While $\mu^{\text{LBAC}}$ has its virtues, it incorrectly predicts zero CT in centrosymmetric excitations. In subsequent sections, the modulus of this metric will be written as $\mu^{\text{LBAC}}$ for simplicity.


Among other metrics unable to describe centrosymmetric excitations, $|\mathbf{r}_D - \mathbf{r}_A|$ (based on detachment and attachment densities)\cite{head1995analysis,plasser2014new} is found to be equivalent to $|\mathbf{r}_e - \mathbf{r}_h|$ (based on the 1-TDM)\cite{plasser2012analysis,plasser2015statistical} for Configuration Interaction Singles (CIS) and TDDFT within the Tamm-Dancoff approximation (TDA).\cite{plasser2015statistical} Also, the modulus of $\mathbf{\mu^{\text{LBAC}}}$ is proven to provide the same result as $|\mathbf{r}_e - \mathbf{r}_h|$ for CIS and TDDFT/TDA in the Supporting Information (Section S7). $\Delta r$ (based upon molecular orbitals)\cite{guido2013metric,guido2014effective} only omits the coupling term between different singly excited Slater determinants when compared to $\mu^{\text{LBAC}}$ (as shown in Section S7 of the Supporting Information). Therefore it is sufficient to select $\mu^{\text{LBAC}}$ from this set of metrics.


The RMS separation of the electron and hole positions $\text{RMS}d_{eh}$ can describe the electron-hole separation in centrosymmetric excitations.\cite{plasser2015statistical}
\begin{equation}
    \text{RMS}d_{eh} = \sqrt{\braket{\chi_\mathrm{exc}(\mathbf{r}_h,\mathbf{r}_e)|(\mathbf{r}_e - \mathbf{r}_h)^2|\chi_\mathrm{exc}(\mathbf{r}_h,\mathbf{r}_e)}}
    \label{eq:RMS-d_eh}
\end{equation}
$\chi_\mathrm{exc}(\mathbf{r}_h,\mathbf{r}_e)$ refers to the exciton wavefunction and it can be represented with the 1-TDM of a quantum chemical excited state calculation.
\begin{equation}
    \chi_\mathrm{exc}(\mathbf{r}_h,\mathbf{r}_e) = N\int \Phi^{\text{GS}}(\mathbf{r}_h,\mathbf{r}_2,...,\mathbf{r}_N)\times\Phi^\mathrm{EX}(\mathbf{r}_e,\mathbf{r}_2,...,\mathbf{r}_N) d\mathbf{r}_2,...,d\mathbf{r}_N
\end{equation}

Another metric that can describe centrosymmetric excitations is $\Lambda$,\cite{peach2008excitation} which calculates and sums over the spatial overlap between molecular orbitals involved in the excitation for each singly excited Slater determinant in CIS and TDDFT/TDA calculations, with values ranging from 0 to 1. However, it has been shown to be an inefficient scale in distinguishing certain short-range CT states from valence states.\cite{guido2013metric} $\phi_S$, which evaluates the overlap between the detachment and attachment density, yields similar results as $\Lambda$, though built from a different theoretical foundation.\cite{etienne2014toward} $\Tilde{d}_{exc}$, which is constructed from the charge transfer number, is a fragmented version of $\text{RMS}d_{eh}$.\cite{mewes2016excitons} Moreover, a modified variant of Le Bahers et al.'s $D_{CT}$ metric, named $D_{CT}^{P}$, has been proposed to effectively characterize centrosymmetric CT states, assessing the CT based on a selective subset of atoms within a molecule\cite{campetella2019quantifying}. However, the determination of such fragment-based metrics inherently relies on fragment selection. This dependency poses challenges for unambiguous comparisons, particularly on centrosymmetric Rydberg excitations. Therefore we think it is sufficient to select only $\text{RMS}d_{eh}$ for comparison with our EMD metric.

In order to make units consistent between the different CT metrics, we choose to define:
\begin{equation}
\mu^{\text{RMS}} = \text{RMS}d_{eh} \cdot 1e
\label{eq:mu_RMS}
\end{equation}
in our comparison. This choice is reasonable because $\text{RMS}d_{eh}$ always calculates the distance between a single electron-hole pair, i.e., one exciton. We can therefore choose the units of all 3 metrics compared in Section~\ref{sec:benchmark} to be \AA $\cdot e$. These results may be converted to the common dipole unit of Debyes based on $1 \ \text{Debye} = 0.208194 \ e \cdot \text{\AA}$.


\subsection{Computational Details} \label{subsec:computational details}

This study utilizes a main dataset comprising 67 single excitations and 3 double excitations from 29 molecules. These excitations can also be categorized into 22 valence excitations, 16 Rydberg excitations, 27 charge transfer excitations, and 5 core excitations based on excitation type. The molecular geometries used for the 5 core excitations are experimental structures from the CCCBDB database,\cite{johnson2022nist} while the others are from the Quest database.\cite{loos2018mountaineering,loos2019reference,loos2020mountaineering,loos2021reference,veril2021questdb}

We perform all the CIS, TDDFT, and OO-DFT calculations using a development version of the Q-Chem quantum chemistry program.\cite{epifanovsky2021software} The excitation space is restricted for core excitations in TDDFT calculations in Section~\ref{subsec:EMD_camb3lyp}, i.e., the electron can only be excited from the core orbital studied to the whole virtual space, to ensure core-valence separation.\cite{cederbaum1980many} The Tamm-Dancoff approximation (TDA) is applied in all TDDFT calculations, with the fact that the impact of using TDA is usually small for $\mu^{\mathrm{EMD}}$ values (see Section S8).\cite{hirata1999time1} The double excitation calculations are performed using the $\Delta$SCF method with the square-gradient minimization (SGM) algorithm.\cite{hait2020excited} Restricted open-shell Kohn-Sham (ROKS)/SGM is applied in calculating core excitations with orbital relaxation in Section~\ref{subsec:double_ex}.\cite{kowalczyk2013excitation, hait2020excited, hait2020highly} The aug-cc-pCVTZ\cite{woon1995gaussian} is employed as the basis set for non-hydrogen atoms in core excitation calculations and aug-cc-pVTZ\cite{dunning1989gaussian,kendall1992electron,woon1993gaussian} is used as the basis set in all other calculations. Excitation energies in Section~\ref{subsec:metric_for_TDDFT} are obtained from our previous benchmark (Ref.~\citenum{liang2022revisiting}) on the performance of TDDFT for electronic excitations. Unless noted otherwise, excitation energies in this paper refer to the energy of vertical excitations. 

For TDDFT and OO-DFT calculations, the numerical quadrature grids are chosen with the radial part treated using the Euler-Maclaurin scheme \cite{murray1993quadrature} and the angular part using the Lebedev scheme.\cite{lebedev1975values}. The XC matrix elements are calculated over a radial grid with 50 points and an angular grid with 194 points for all atoms. As shown in the Supporting Information (Section S1), this level of quadrature grid is large enough to accurately determine excitation energies and electronic densities in our data set, with only a small RMSE relative to reference calculations using SG-3.\cite{dasgupta2017standard} The convergence threshold of the SCF iteration is $10^{-7}$ Hartree and the integration threshold is $10^{-11}$. IQmol\cite{gilbert2012iqmol} is used for the visualization of molecular orbitals and natural transition orbitals.

Transportation simplex algorithm\cite{rubner1998code} is applied to get the EMD of charge distributions. The size of the key grid points is 19 (radial) $\times$ 26 (angular) for each atom, ensuring accuracy. The code for calculating $\mu^{\text{EMD}}$ is provided through Github at 
\newline
https://github.com/zhewang233/ChargeEMD.git

\section{Results and Discussion} \label{sec:benchmark}
\subsection{Comparison of $\mu^{\text{EMD}}$, $\mu^{\text{LBAC}}$, and $\mu^{\text{RMS}}$ in TDDFT calculations}
\label{subsec:EMD_camb3lyp}

\begin{figure}[ht!]
    \centering
    \includegraphics[width=\textwidth]{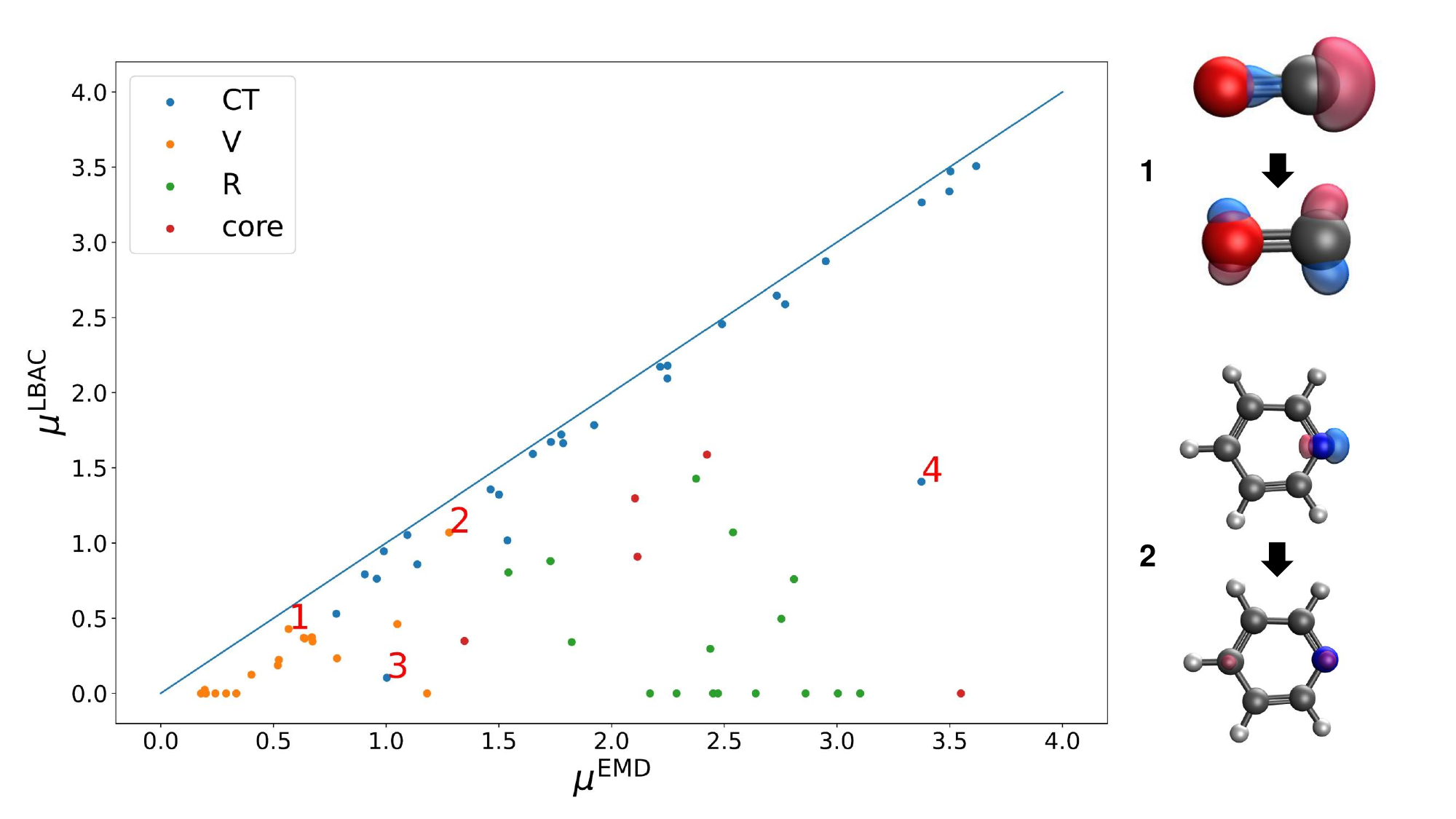}
    \caption{Theoretical metrics for the extent of charge transfer are compared for four types of excitations. The metrics are generated via TDDFT calculations at the CAM-B3LYP level. The blue line represents a slope of 1, indicating a one-to-one relationship. The natural transition orbital pairs with the highest weight (denoted as HONTO and LUNTO respectively here and below)} of two unidirectional valence excitations are shown. The iso-value for the NTO surface is set to be 0.5 \AA$^{-3}$.
    \label{fig:EMD_vs_dCT}
\end{figure}

For all results presented in this subsection, we calculate the CT metrics from amplitudes evaluated via TDDFT (in the Tamm-Dancoff approximation) using the CAM-B3LYP\cite{yanai2004new} functional. CAM-B3LYP is a range-separated hybrid (RSH) functional that can manage the challenging charge transfer excitations quite effectively.\cite{liang2022revisiting}

Figure~\ref{fig:EMD_vs_dCT} is a parity plot of the values of $\mu^{\text{LBAC}}$ and $\mu^{\text{EMD}}$ for each of the 67 single excitations in our dataset. Each data point is color-coded by the class of excitation: CT, valence, Rydberg, and core. Focusing first on the cluster of points close to the parity line, it is evident that both metrics quantitatively agree on the extent of charge separation in unidirectional valence excitations such as excitations in carbon monoxide (labeled point 1) and pyridine (labeled point 2) and the charge transfer excitations (blue points). Given the independent nature of $\mu^{\text{LBAC}}$ and $\mu^{\text{EMD}}$, it is encouraging that they produce similar results for these unidirectional excitations. Nonetheless, it is noteworthy that the magnitude of  $\mu^{\text{EMD}}$ is always greater than that of $\mu^{\text{LBAC}}$, reflecting that fact that electron density is actually rearranging in three dimensions, and, for example, the electron rearrangement perpendicular to the bond axis in CO is centrosymmetric, which cannot be captured by $\mu^{\text{LBAC}}$.

For Rydberg and core excitations, $\mu^{\text{LBAC}}$ deviates much more from the linear relationship and is much smaller than $\mu^{\text{EMD}}$. This is due to the fact that in Rydberg and core excitations, the electron cloud movement is often centrosymmetric (or pseudo-centrosymmetric) and the description of the distance between the electron and hole centers alone ($\mu^{\text{LBAC}}$) is insufficient to reflect the degree of charge separation in such excitations. In contrast, $\mu^{\text{EMD}}$ can describe the movement of the charge distribution as a whole and therefore produces meaningful values in these (pseudo-)centrosymmetric excitations. For example, $\mu^{\mathrm{EMD}}$ could capture the charge transferred from the center to both sides in the A-D-A type molecules with the electron-donating group in the middle and electron-withdrawing groups on both sides, which is the shortest distance for transportation. From another perspective, the extent of centrosymmetry of a given excitation can be distinguished clearly from a comparison between $\mu^{\text{LBAC}}$ and $\mu^{\text{EMD}}$. As shown in Figure~\ref{fig:EMD_vs_dCT}, some CT excitations (points 3 and 4) show a relatively smaller value of $\mu^{\text{LBAC}}$ and a relatively larger value of $\mu^{\text{EMD}}$ due to partial Rydberg character in these two excitations (see details in Figure S1).



\begin{figure}[ht!]
    \centering
    \includegraphics[width=0.80\textwidth]{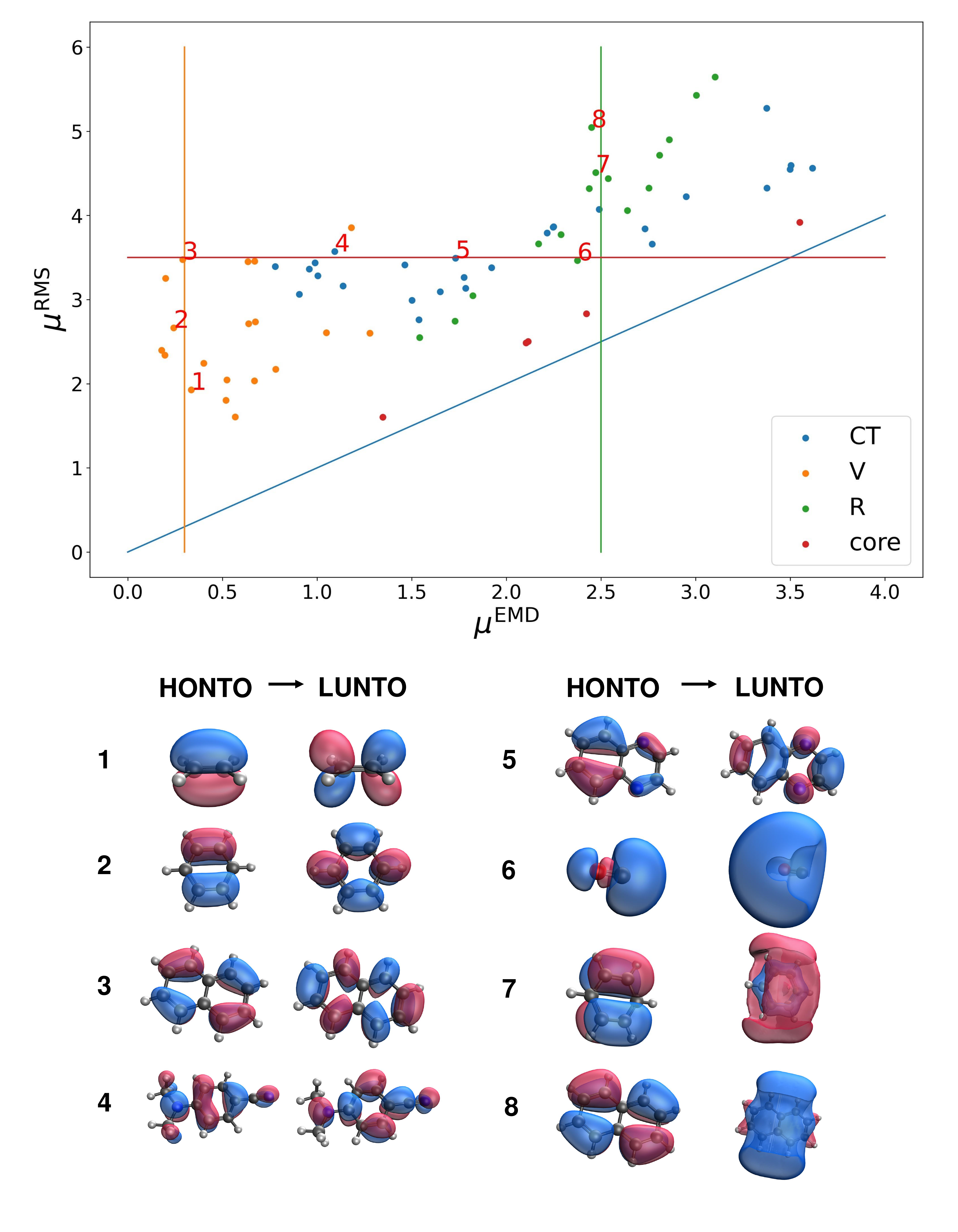}
    \caption{Comparison between $\mu^{\text{RMS}}$ and $\mu^{\text{EMD}}$. The HONTOs and LUNTOs are shown for some excitations. The iso-value for the NTO surface is set to be 0.1 \AA$^{-3}$.}
    \label{fig:EMD_vs_dexc}
\end{figure}

Figure~\ref{fig:EMD_vs_dexc} assesses the extent of correlation between the values of $\mu^{\text{RMS}}$ (see Equation \ref{eq:mu_RMS}) and $\mu^{\text{EMD}}$ for the 67 single excitations in our dataset. The root-mean-square feature of $\mu^{\text{RMS}}$ (Equation \ref{eq:RMS-d_eh}) inherently grants more significance to extended electron-hole distances, resulting in larger estimates than $\mu^{\text{EMD}}$. This is especially apparent when dealing with excitations involving significant electron or hole sizes. For example, the natural transition orbitals (NTOs) for points 1-3 (near the yellow vertical line of Figure~\ref{fig:EMD_vs_dexc}) suggest these 3 transitions are all valence $\pi \to \pi^*$ excitations. For these 3 specific excitations, $\mu^{\text{EMD}}$ yields comparable results, whereas $\mu^{\text{RMS}}$ values show a correlation with molecular size ($1<2<3$).  This same trend is also visible when comparing 3 Rydberg states close to the vertical green line (points 6-8), for which $6<7<8$.

Turning our attention to points 3 and 6 (along the horizontal red line of Figure~\ref{fig:EMD_vs_dexc}), we see that $\mu^{\text{EMD}}$ exhibits a small value for the $\pi \to \pi^*$ transition in naphthalene (3) and predicts a more substantial value for a Rydberg excitation in carbon monoxide (6). These $\mu^{\text{EMD}}$ results seem very consistent with the different changes in electronic density in these two excitations as shown in the NTOs. In contrast,  $\mu^{\text{RMS}}$ yields similar values for these two distinct excitations, 3 and 6. 

From these observations, it seems clear that $\mu^{\text{EMD}}$ is responsive to the nature of the density change, and is relatively insensitive to molecular size (or more specifically, the sizes of the electron and hole). The same cannot be said for $\mu^{\text{RMS}}$. Therefore,  $\mu^{\text{EMD}}$ can clarify the distinction between valence excitations and excitations with larger density changes (such as Rydberg excitations and long-range CT excitations) more clearly than $\mu^{\text{RMS}}$, giving trends that are more in line with visual inspection of the NTOs.


\subsection{Utilization of $\mu^{\text{EMD}}$ in OO-DFT calculations}
\label{subsec:double_ex}

\begin{figure}[ht!]
    \centering
    \includegraphics[width=\textwidth]{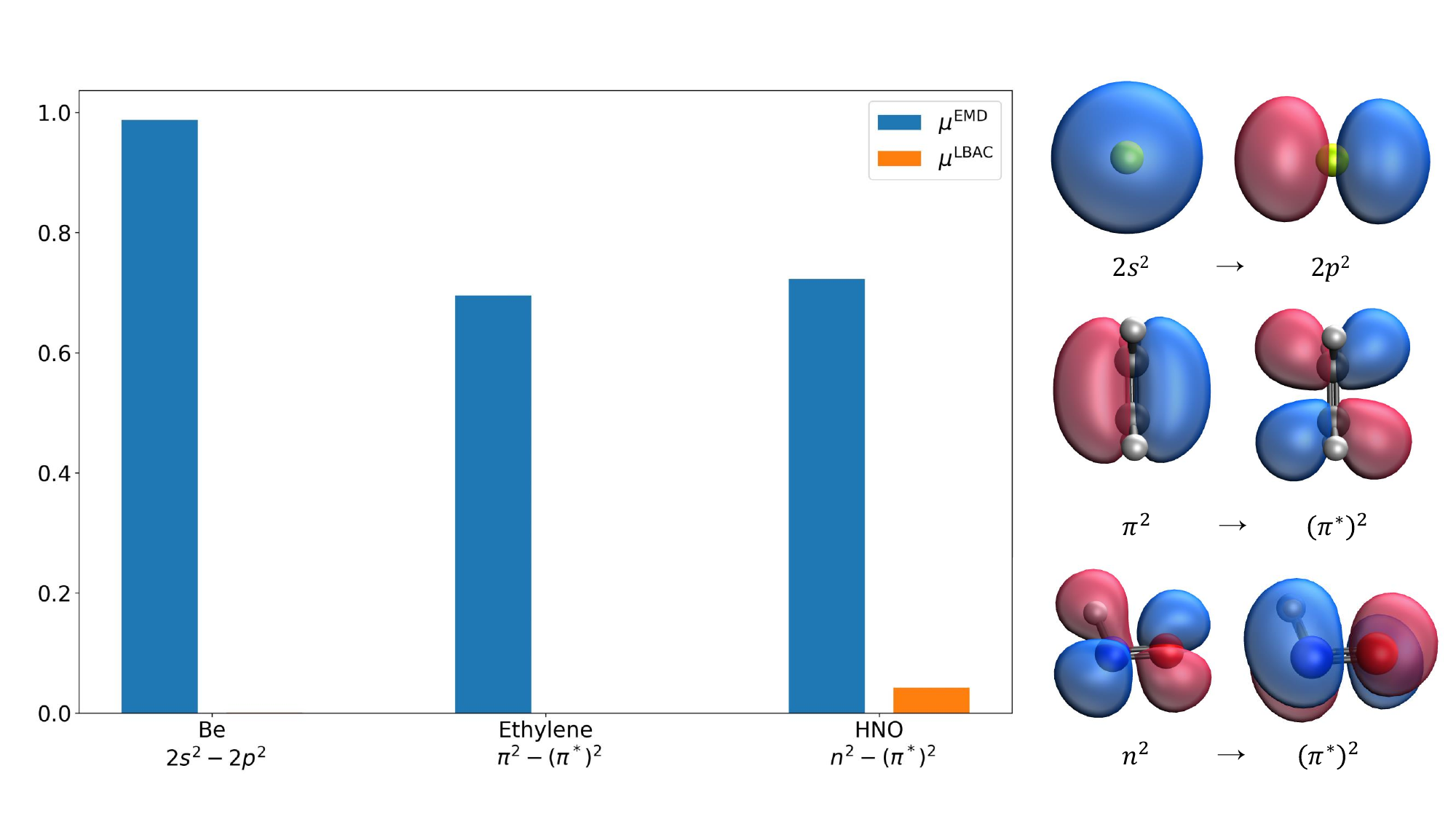}
    \caption{Comparison between $\mu^{\text{EMD}}$ and $\mu^{\text{LBAC}}$ calculated from densities via $\Delta$SCF/aug-cc-pVTZ calculations using the CAM-B3LYP functional with the SGM optimization method, on three simple double excitations. The canonical orbitals of these excitations are displayed to the right. From top to bottom, they are $2s^2 \to 2p^2$ in Be, $\pi^2 \to (\pi^*)^2$ in ethylene, and $n^2 \to (\pi^*)^2$ in HNO.}
    \label{fig:double_ex}
\end{figure}

Due to the adiabatic local-density approximation (ALDA), it is difficult for linear-response (LR) TDDFT to model double and higher excitation states.\cite{hait2021orbital,liang2022revisiting} Besides, TDDFT struggles to capture the relaxation effect in certain double and core excitations because the same ground state orbitals are used to represent both ground and excited states. Since $\mu^{\text{EMD}}$ is fully based on the electronic densities, it can be easily applied to analyze results obtained with other excited state theories such as OO-DFT\cite{hait2021orbital} that are more appropriate for excitations that pose challenges for TDDFT. 

In double excitations, the 1-TDM cannot capture the 2-electron excited configurations, thereby strongly limiting the usefulness of 1-TDM-based metrics such as $\mu^{\text{RMS}}$. We selected three simple double excitations and Figure~\ref{fig:double_ex} displays their $\mu^{\text{EMD}}$ and $\mu^{\text{LBAC}}$ values, calculated from OO-DFT densities using the CAM-B3LYP functional. Evidently, $\mu^{\text{EMD}}$ gives reasonable values for centrosymmetric double excitations (e.g., $2s^2 \to 2p^2$ in Be and $\pi^2 \to (\pi^*)^2$ in ethylene) and double excitations that contain significant density rotations (e.g., $n^2 \to (\pi^*)^2$ in HNO), whereas $\mu^{\text{LBAC}}$ falls short.

We present the $d^{\text{EMD}}$, $q^{\text{CT}}$, and $\mu^{\text{EMD}}$ values for the single and double excitations from $\pi$ to $\pi^*$ in ethylene in Table~\ref{tab:single_vs_double}. As expected, $\mu^{\text{EMD}}$ for the  double excitation is nearly twice that of the single excitation, primarily because $q^{\text{CT}}$ doubles in value while $d^{\text{EMD}}$ remains virtually unchanged. This suggests that these single and double excitations have a similar spatial extent of charge transfer, with only minor orbital relaxation.

\begin{table}[ht!]
\caption{A comparison between single and double excitation from $\pi$ to $\pi^*$ in ethylene.}
\begin{tabular}{lccc}
\hline
Excitation type & \multicolumn{1}{l}{$d^{\text{EMD}}$} & \multicolumn{1}{l}{$q^{\text{CT}}$} & \multicolumn{1}{l}{$\mu^{\text{EMD}}$} \\ \hline
Single & 1.1845  & 0.2829 & 0.3351 \\
Double & 1.1797 & 0.5895 & 0.6954 \\ \hline
\end{tabular}
\label{tab:single_vs_double}
\end{table}

\begin{figure}[ht!]
    \centering
    \includegraphics[width=0.7\textwidth]{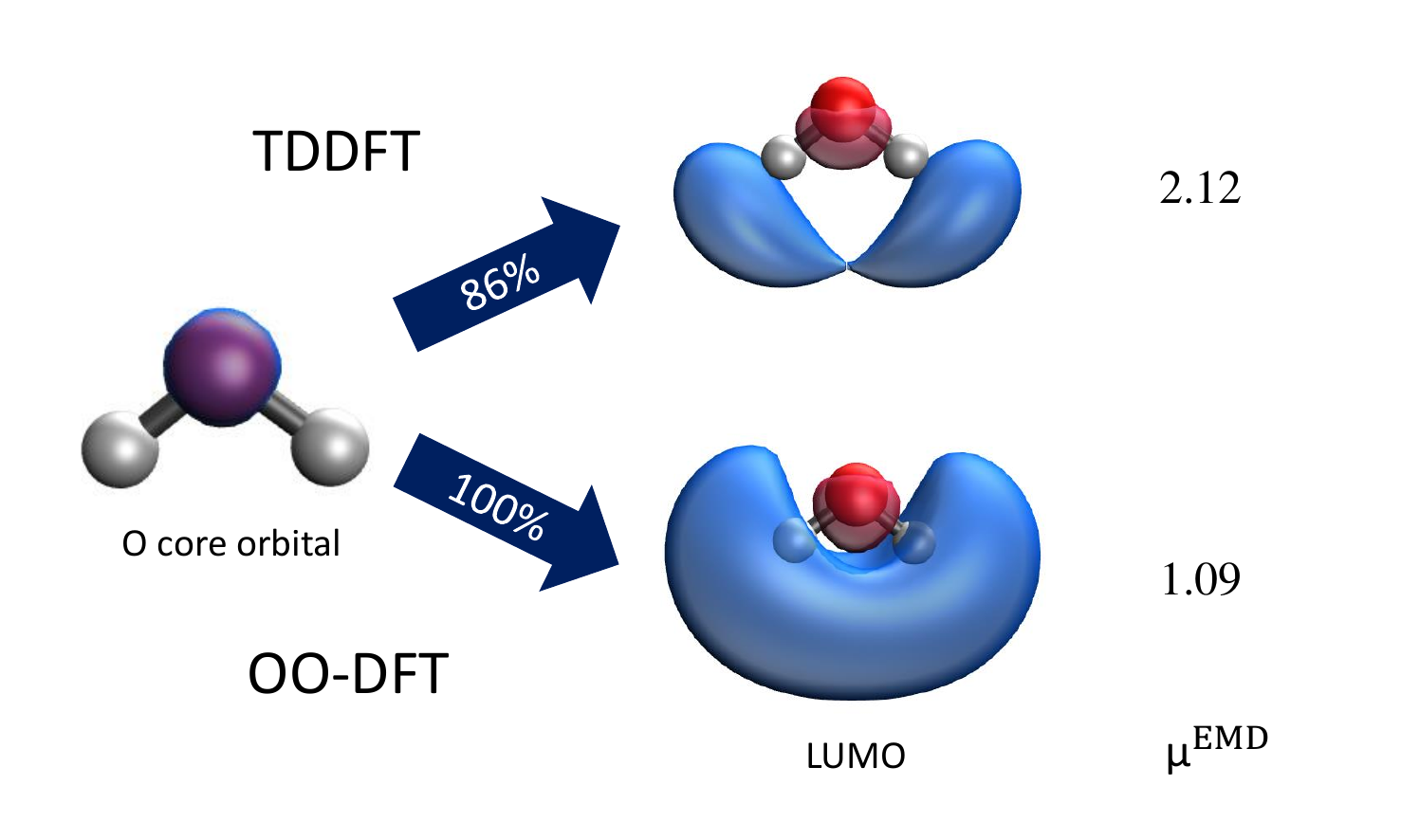}
    \caption{Comparison between TDDFT and OO-DFT in the excitation of water from the O core orbital to the LUMO. In TDDFT calculation, the singlet excitation from the Oxygen core orbital to the vacant space with the lowest excitation energy is investigated, in which the excitation to LUMO contributes 86$\%$ and is considered to be the main excitation. In OO-DFT, the full excitation from the Oxygen core orbital to LUMO is studied. The iso-value for the molecular orbital surface is set to be 0.1 \AA$^{-3}$.}
    \label{fig:core_oodft_vs_tddft}
\end{figure}

$\mu^{\text{EMD}}$ can also quantitatively capture the much larger orbital relaxation effect in core excitations when utilized with OO-DFT.\cite{hait2020highly} This can be demonstrated by investigating a core excitation in the water molecule (Figure~\ref{fig:core_oodft_vs_tddft}). In the core excitation, the outer valence orbitals of water are expected to contract inward due to the promotion of the core electron, and thus a decrease in the screening of the nuclear charge. Compared to TDDFT, the OODFT calculation provides a significantly smaller $\mu^{\text{EMD}}$ value, quantifying the extent of orbital relaxation in this core excitation relative to TDDFT.

Considering the inability of $\mu^{\text{LBAC}}$ to measure the extent of density rearrangement in (pseudo-)centrosymmetric excitations and the unsuitability of 1-TDM-based metrics for double excitations, $\mu^{\text{EMD}}$ appears to be the best choice for integration with OO-DFT methods to study the double and core excitations. At the same time, there is no substitute for examining the NTOs or attachment and detachment densities in order to reliably assign the character of a state. 

\subsection{Influence of different functionals on $\mu^{\text{EMD}}$ in TDDFT calculations}
\label{subsec:EMDs}

\begin{figure}[ht!]
    \centering
    \includegraphics[width=\textwidth]{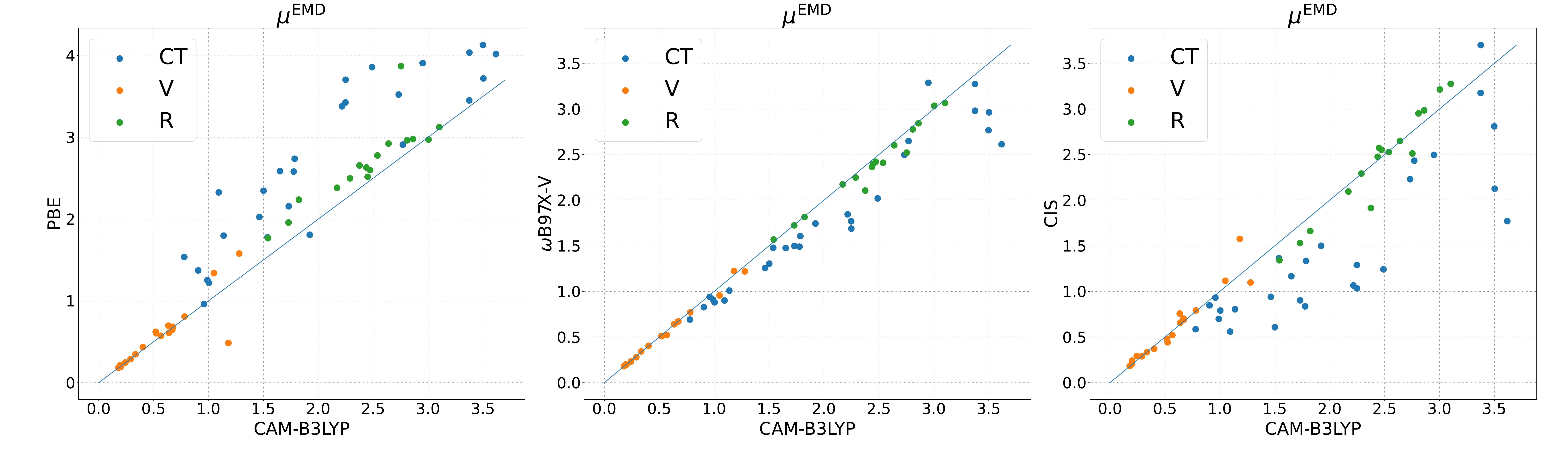}
    \caption{Comparison of $\mu^{\text{EMD}}$s based on density calculated from theoretical methods with different Hartree-Fock (HF) exchange components (PBE, CAM-B3LYP, $\omega$B97X-V, and CIS). The blue line represents a slope of 1, indicating a one-to-one relationship.}
    \label{fig:EMD_functionals}
\end{figure}

Different theories, and in particular different density functionals within TDDFT, typically yield different excitation energies and different electronic densities, resulting in different metric values. It is possible that comparing $\mu^{\text{EMD}}$ values for different functionals or methods on a common dataset of excitations could reveal useful information about systematic differences in the density changes between different functionals, as a result of their different XC treatment. Such differences include the fraction of exact [Hartree-Fock (HF)] exchange, as well as the different treatment of semi-local exchange and correlation. 

To explore this possibility, Figure~\ref{fig:EMD_functionals} provides a detailed comparison of the $\mu^{\text{EMD}}$ values derived from electronic densities generated by CIS and TDDFT calculations utilizing PBE\cite{perdew1996generalized}, CAM-B3LYP, and $\omega$B97X-V\cite{mardirossian2014omegab97x} functionals, respectively. For most valence and some Rydberg excitations, similar $\mu^{\text{EMD}}$ results are obtained for all 4 theories, indicating that the hole/electron distributions are relatively insensitive to the differences between these methods. However, in most charge transfer and some Rydberg excitations, a decrease in predicted $\mu^{\text{EMD}}$ values is observed with an increasing fraction of exact exchange, from PBE to CIS, which is associated with an increase in excitation energy.  The trend in $\mu^{\text{EMD}}$ values implies a decrease in the extent of electron-hole separation with an increasing fraction of exact exchange. A comparison between functionals using other metrics also supports this conclusion (Figures S2 and S3). Similar outcomes were observed in Le Bahers et al.'s study of donor-acceptor systems (dyads), where PBE and PBE0 indicated more pronounced charge-hole separation compared to LC-PBE and CIS.\cite{le2011qualitative} 

Since we did not find theoretical explanations for this phenomenon based on TDDFT theory in previous papers, we offer one possible explanation which can be derived from the general expression of the $\mathbf{A}$ matrix from the LR-TDDFT equation (shown for a global hybrid functional):
\begin{equation}
    \mathbf{A}_{ia,jb} = \delta_{i,j}\delta_{a,b}(\epsilon_a - \epsilon_i) + (ia|jb) - c_{\text{HF}}(ij|ab) + (1-c_{\text{HF}})(ia|f_{xc}|jb).
\end{equation}
The four terms refer to the difference in one-particle orbital energies, the response of the Coulomb potential, the response of the exact exchange potential, and the response of the chosen XC potential, respectively. In a long-range charge transfer state where the donor and acceptor orbital overlap is minimal, the second and fourth terms make minor contributions to the $\mathbf{A}$ matrix. The third term becomes dominant in such situations, accounting for the electrostatic attraction between the created hole (orbital $i$, $j$) and the electron (orbital $a$, $b$).\cite{dreuw2005single} When the functional has more non-local HF exchange, the third term becomes more significant, lowering the energy of excited states with more substantial electrostatic attraction between holes and electrons, resulting in low-lying states with more compact electron-hole pairs.  


In addition, the EMD metric can serve as a robust indicator to aid in the assignment of excited states in TDDFT calculations when using different functionals.  It is shown to be useful for correcting some mistakenly assigned excited states in previous benchmark work (See Supporting information).\cite{liang2022revisiting}

\subsection{Potential diagnostic tool for performance of functionals in TDDFT calculations}
\label{subsec:metric_for_TDDFT}

\begin{figure}[ht!]
    \centering
    \includegraphics[width=\textwidth]{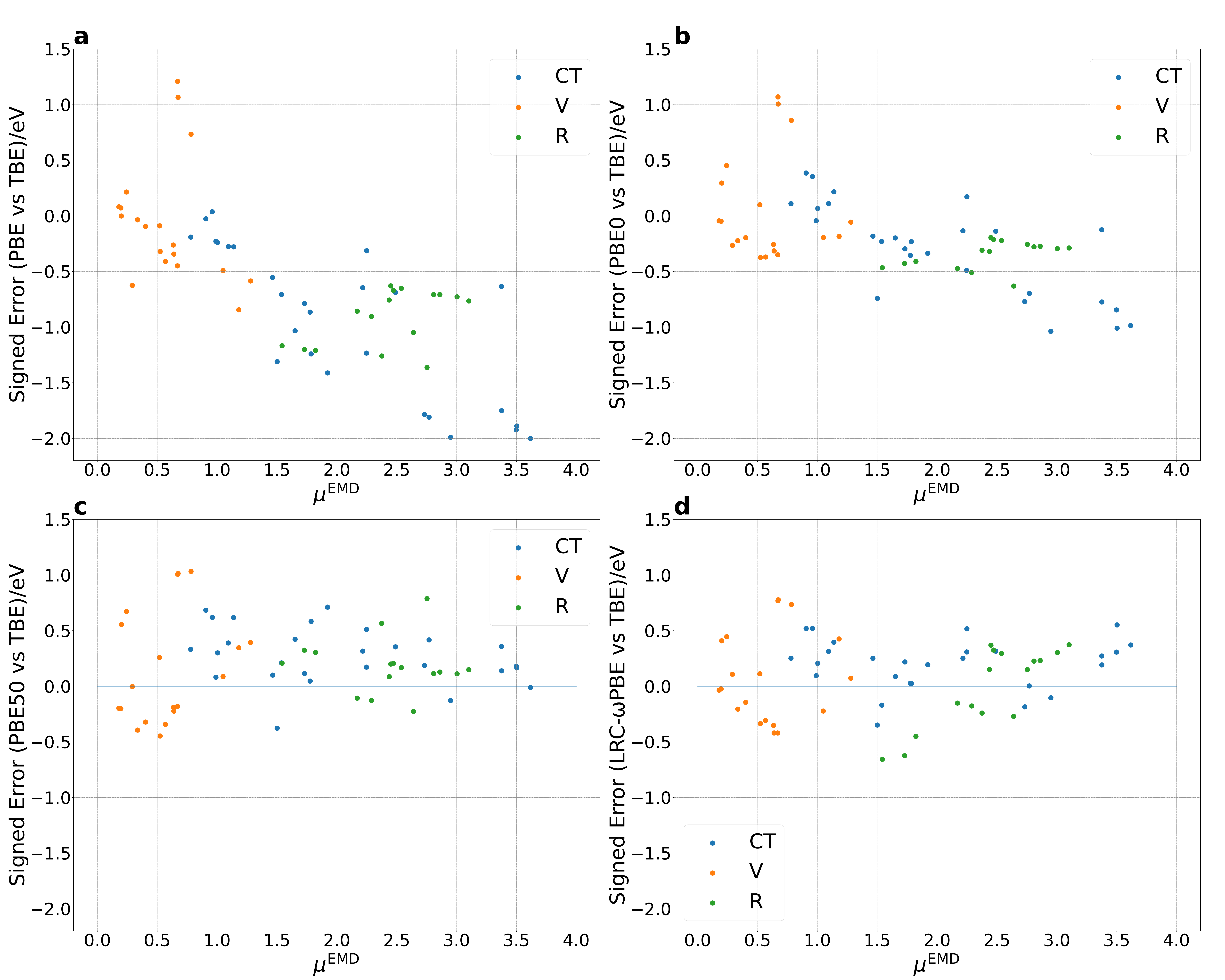}
    \caption{Comparison of the performance of (a)PBE, (b)PBE0, (c)PBE50, and (d)LRC-$\omega$PBE in TDDFT excitation energies across valence, Rydberg, and charge transfer excitations. The presented signed errors in excitation energy are from Ref.~\citenum{liang2022revisiting}. $\mu^{\text{EMD}}$ is computed from densities generated in the TDDFT calculation using the CAM-B3LYP functional.}
    \label{fig:funcs_vs_TBE}
\end{figure}

LR-TDDFT is widely used in computing single electron excitation energies due to its computational efficiency and accuracy. However, pure local XC functionals are often unable to accurately predict the excitation energies for certain states with substantial density differences from the ground state, such as charge transfer, Rydberg, and core excitations.\cite{liang2022revisiting} The EMD metric offers a means of quantifying the density change in the excitation, suggesting that $\mu^{\text{EMD}}$ could also indicate potentially poor quality results. For instance, $\mu^{\text{EMD}}$ may be able to flag failures of pure local XC functionals that are associated with significant density changes between the ground and excited states.

To explore this possibility, we have compared the performance of several functionals, namely PBE,\cite{perdew1996generalized} PBE0,\cite{adamo1999toward} PBE50,\cite{bernard2012general} and LRC-$\omega$PBE,\cite{rohrdanz2008simultaneous} for TDDFT excitation energies across valence, Rydberg, and charge transfer excitations. This comparison is presented in Figure~\ref{fig:funcs_vs_TBE}, where the densities generated from the TDDFT calculation using the CAM-B3LYP functional are used to evaluate $\mu^{\text{EMD}}$.

As a semi-local XC functional, PBE achieves acceptable accuracy for most valence and short-range charge-transfer excitations, where $\mu^{\text{EMD}}$ is below 1.0. However, as $\mu^{\text{EMD}}$ increases, PBE's underestimation of excitation energies becomes more pronounced, particularly for long-range charge transfer and Rydberg excitations. The introduction of non-local exact HF exchange in hybrid functionals, exemplified by PBE0, results in a less negative slope with increasing $\mu^{\text{EMD}}$, indicating the mitigation of systematic errors (self-interaction errors).  With increasing fraction of exact exchange, PBE50 achieves better performance for most Rydberg and long-range CT excitations although it tends to overestimate these excitation energies. For RSH functionals with high fractions of non-local exact exchange for long-range electronic interactions, taking LRC-$\omega$PBE as an example, the error is small for those difficult excitations. Earlier research has highlighted diverse metrics as valuable diagnostic tools for assessing the performance of TD-DFT calculations\cite{adamo2015exploring, peach2008excitation, etienne2014toward, guido2013metric}. Notably, $\Lambda$\cite{peach2008excitation} and $\phi_S$\cite{etienne2014toward} metrics, which can deal with centrosymmetric molecules, provide similar inferences for PBE, B3LYP, and CAM-B3LYP functionals. However, the range of our $\mu^{\mathrm{EMD}}$ metric, enlarged by $d^{\mathrm{EMD}}$ spanning from 0 to infinity, offers a more lucid distinction between valence excitations and short-range CT excitations. A detailed explanation for this phenomenon is given in Section S5. 


Through these observations, we suggest that $\mu^{\text{EMD}}$ can serve as a diagnostic tool for predicting the potential failure of pure and certain global hybrid XC functionals. Considering that $\mu^{\text{EMD}}$ is minimally affected by molecular size (as discussed in section~\ref{subsec:EMD_camb3lyp}), we conclude that a  $\mu^{\text{EMD}}$ value of less than 1.0 indicates little spatial transfer of density in the corresponding excitations, such as in valence and short-range CT excitations. Pure XC functionals usually perform adequately for excitation energies in this class. However, when $\mu^{\text{EMD}}$ exceeds 1.0, the excitations are associated with density differences that show significant spatial transfer, leading to potential failures of pure XC and hybrid XC functionals with a small exact exchange component in TDDFT calculations. This error further escalates as $\mu^{\text{EMD}}$ grows. The performance of more functionals are displayed in Figure S7.

\section{Conclusions} \label{sec:conclusion}

We have presented a novel theoretical metric, $\mu^{\text{EMD}}$, to characterize the density change (or extent of charge transfer) in electronic excitations. $\mu^{\text{EMD}}$ is an adaptation of the earth mover's distance (EMD) to the discretized difference density associated with an electronic excitation. This new metric is consistent with $\mu^{\text{LBAC}}$ for unidirectional excitations, but it resolves the limitation of $\mu^{\text{LBAC}}$ for describing (pseudo-)centrosymmetric excitations. Compared to 1-TDM-based metrics like $\mu^{\text{RMS}}$, $\mu^{\text{EMD}}$ likewise shows advantages. It is not much influenced by molecular size  (or more precisely, the electron and hole sizes). It can also be readily implemented for a range of excitation methods beyond TDDFT, such as OO-DFT, facilitating the characterization of challenging cases like double and core excitations, and illustrating the effect of orbital relaxation. Therefore, $\mu^{\text{EMD}}$ will serve as a useful scalar to compactly characterize the extent of charge transfer.

Furthermore, our study suggests that $\mu^{\text{EMD}}$ can be utilized as a diagnostic tool to signal potentially poor numerical results for excitation energies using pure and certain global hybrid XC functionals. This is because such poor results correlate well with the extent of charge transfer including that associated with Rydberg and core excitations which may be (pseudo-)centrosymmetric. In addition, we have observed that the calculated $\mu^{\text{EMD}}$ values for long-range charge transfer excitations will decrease when using functionals with larger fractions of exact exchange. 

It is worth noting that the present calculation time of $\mu^{\text{EMD}}$ is much higher than that of other metrics. However, when juxtaposed with TDDFT calculations, the time taken remains within an acceptable range. Going forward, it is potential to refine the algorithms used to solved the EMD optimization problem, possibly by leveraging modern network simplex algorithms or improving the strategies for grid point selection. Such endeavors could substantially mitigate the computational demands of $\mu^{\text{EMD}}$.

\section*{Supporting Information}

Additional information and figures (SI.pdf)

raw\_data.xlsx

\begin{acknowledgement}
This work was supported by the Director, Office of Science, Office of Basic Energy Sciences, of the U.S. Department of Energy through the Gas Phase Chemical Physics Program, under Contract No. DE-AC02-05CH11231. This research used computational resources of the National Energy Research Scientific Computing Center, a DOE Office of Science User Facility supported by the Office of Science of the U.S. Department of Energy under Contract No. DE-AC02-05CH11231.
\end{acknowledgement}

\bibliography{a-main}
\end{document}